\pdfoutput=1

\documentclass[aps,prl,showpacs,twocolumn,superscriptaddress,showpacs,groupedaddress,nofootinbib]{revtex4-1}  
\usepackage{color}
\usepackage{graphicx} 
\usepackage{dcolumn} 
\usepackage{bm}       
\usepackage{amssymb}   
\usepackage{amsmath,amssymb,amsthm,mathrsfs,amsfonts,dsfont} 
\usepackage[colorlinks=true,citecolor=black, linkcolor=black]{hyperref}
\usepackage[dvipsnames]{xcolor}

\definecolor{darkblue}{rgb}{0.0, 0.0, 0.55}

\newcommand{\myhref}[1]{\href{https://arxiv.org/abs/#1}{{\color{black} #1}}}

\def\be{\begin{eqnarray}}
\def\ee{\end{eqnarray}}
\def \bea {\begin{eqnarray}}
\def \eea {\end{eqnarray}}

\def \nn {\nonumber}
\def \la {\langle}
\def \ra {\rangle}
\def \del {\partial}
\def \a {\alpha}

\def \b {\beta}

\def \g {\gamma}

\def \m {\mu}
\def \n {\nu}

\def\frac#1#2{{#1\over #2}}

\begin{document}

\title{Stress-Tensor Commutator in CFT near the Lightcone}
\author{Kuo-Wei Huang \vspace{0.1cm}}
\affiliation{
Department of Physics, Boston University, \\
Commonwealth Avenue, 
Boston, Massachusetts 02215 
}
\date{\today}

\fontsize{10pt}{12pt}\selectfont

\begin{abstract}

Starting with the general stress-tensor commutation relations consistent with the 
Poincar\'e algebra in local quantum field theory, we impose the tracelessness condition 
and focus on the dominating contributions in the lightcone limit.  
It is shown that, under a certain assumption on the Schwinger term, 
a Virasoro-algebra-like structure emerges near the lightcone in $d>2$ conformal field theories. 

\end{abstract}
\maketitle

As the universal central extension of the Witt algebra, the 
existence of the Virasoro algebra \cite{V70} 
plays a crucial role in mathematics and theoretical physics, 
particularly of deep importance in conformal field theory (CFT).
It is, however, a special luxury one has in two-dimensional spacetime.  
In higher dimensions, where the conformal group is finite dimensional, 
Virasoro-algebra related techniques employed in understanding 
$d=2$ CFT  become generally invalid. 
Nevertheless, one may still ask the following question: 
does 
an effective similar structure exist in higher-dimensional CFTs  
that can be used to control the CFT data within 
a certain {\it subspace} (i.e. a subsector of the full parameter space)? 

As the form of two commuting copies, the Virasoro algebra can  
be expressed as the stress-tensor commutation relation: 
\be
\label{V1}
&&- i  [T^{++}(x^-), T^{++}(x'^-)]\nn\\
&=&  4 \Big( T^{++}(x^-)+ T^{++}(x'^-) \Big) \del_- \delta(x^--x'^-)\nn\\
&&~~~~~~~~~~~~~~~~~~~~~~~~~~~~~~- {2 c\over 3 \pi}\del^3_- \delta(x^--x'^-)\ ,
\ee 
with a similar expression for $[T^{--}(x^+), T^{--}(x'^+)]$. 
We denote  
$(x^+, x^-)= (t+y, t - y)$ with $(x^0, x^1)= (t, y)$,
and  
$(T^{++}(x^-),T^{--}(x^+)) = -2 (T^0_0- T^0_1 ,~ T^0_0+ T^0_1)$ in 
Minkowski metric $\eta_{\mu\nu}= {\rm diag}(-1,1)$. 
The central-extension part containing the 
central charge $c$ arises from the quantum anomaly. 
While the tracelessness condition in $d=2$
allows one to replace the purely spatial-component of 
the stress tensor, $T_{11}$, with $T_{00}$, independent 
spatial-components  appear in $d>2$,
and, in general, there is no stress-tensor algebra in higher-dimensions.  
In this work, we make an initial attempt, starting from the most 
general stress-tensor commutation relations in Lorentz invariant, local 
quantum field theory (QFT) \cite{JS1962, JS1963, BD1967}, to search 
for a possible Virasoro-like structure in higher-dimensional CFTs.
In particular, while the tracelessness constraint must be imposed, 
we would like to see under what additional conditions
an effective Virasoro-like algebra may emerge.

What clue do we have? 
The AdS/CFT correspondence 
\cite{Maldacena:1997re,Witten:1998qj,Gubser:1998bc} 
provides an interesting hint toward this direction.
In a recent work \cite{Fitzpatrick:2019zqz}, it was found 
that the operator product expansion
(OPE) coefficients of the multi stress-tensor 
conformal blocks of a scalar four-point function in 
a large class  of $d>2$ CFTs 
are universal in the {\it lowest-twist} subspace.  
(The twist of an operator is its dimension minus its spin.)
These isolated OPE coefficients are universally fixed by the 
dimensions of the light and heavy scalar operators, and 
the central charge $C_T$, the coefficient of the stress-tensor 
two-point function. In $d=2$, the Virasoro algebra dictates 
all the related structures. 
While additional assumptions were made in the gravitational 
computation considered in \cite{Fitzpatrick:2019zqz}, such 
as having a large $C_T$ and ignoring additional bulk matters, 
it is tempting to ask if the lowest-twist limit is essentially 
sufficient in a more general analysis in $d>2$ CFTs.
As the lowest-twist limit corresponds to the lightcone limit, where 
operators in a correlator approach each other's lightcone, we are therefore 
motivated to consider the CFT stress-tensor  commutators  near the lightcone.

In the next section, we first review the general stress-tensor equal-time 
commutation relations in QFT, based on earlier works \cite{JS1962, JS1963, BD1967}. 
The tracelessness condition and the lightcone limit shall be imposed in a later section. 
The main result is to obtain an  effective lightcone stress-tensor commutator in CFT. 
By effective, we mean that  the lightcone limit is taken when  
stress tensors are inserted in a correlation function. 
In this case, the purely lightcone-component of the 
stress tensor, denoted as $\widetilde T^{++}$ below, dominates the contributions.
A crucial point is that, in such an effective lightcone limit, one 
avoids the commutator with purely spatial-components, i.e. $[T_{ij}(x), T_{mn}(x')]$, 
whose form cannot be determined by Poincar\'e symmetry or 
conformal invariance and thus is generally model-dependent in $d>2$.
(The central charge $C_T$ is also model-dependent, but we say a quantity has
a universal meaning if all the model-dependent data can be absorbed into $C_T$.)  
The resulting effective lightcone commutator will 
have a non-extension part and also a Schwinger term. 
The  non-extension part formally looks the same as that 
in the $d=2$ Virasoro algebra.
The Schwinger term  in general dimensions can be   
related to the central charge  $C_T$.
Some subtleties of the Schwinger term will also be discussed. 

\vskip 0.1in
\noindent
{\it Stress-Tensor Commutation Relations in QFT.---} 
Here we first review the stress-tensor commutators in Lorentz invariant, local QFT 
(see \cite{BD1967} and \cite{JS1962, JS1963} for more discussions).  
Denote a classical action  $S$ embedded in a curved background and   
write the curved-space stress tensor as 
\be
C^{\mu\nu}(x)=2 { \delta S\over \delta g_{\mu\nu}(x)}    \ .
\ee  
A factor $\sqrt{-g}$ is normally factored out from $C_{\mu\nu}$ as the  
conventional stress tensor, but we adopt the above notation for later convenience. 
Eventually, we will be interested in the commutation relations of 
the flat-space stress tensor, denoted as $T^{\mu\nu}(x)$, in metric 
$ds^2= -dt^2 +\delta_{ij} dx^i dx^j$.   
Below, we denote 
$\langle C^{\mu\nu} (x)\ra = -2 i { \delta Z \over \delta g_{\mu\nu}(x)}$, where 
$Z$ 
is the partition function. 
The starting point is to consider the variation of the conservation equation,
\be
\label{conservation}
0=\langle C^{\mu\nu}_{~~;\nu}(x) \ra=\langle C^{\mu\nu}_{~~,\nu}(x)\ra 
+ \langle\Gamma^\m_{\a\b}(x) C^{\a\b}(x) \ra  \ ,
\ee 
with respect to an external  metric.  
Varying the first term on the right-hand side of \eqref{conservation} yields   
\be
2 \del_\nu {\delta \langle  C^{\mu\nu}(x) \ra \over \delta g_{\lambda\rho}(x')}
= \del_\nu \Big( i  \langle C^{\mu\nu} (x) C^{\lambda\rho} (x')\ra_+  +  2   
 \langle  {\delta   C^{\mu\nu} (x) \over  \delta g_{\lambda\rho}(x')} \ra   \Big)\ , \nn
\ee 
where $ \langle C_{\mu\nu} (x) C_{\lambda\rho} (x')\ra_+$ is the time-ordered Lorentzian 
stress-tensor two-point correlator, while varying the second term on 
the right-hand side of \eqref{conservation} gives 
\be
\label{two}
&&2 {\delta \langle\Gamma^\m_{\a\b}(x) C^{\a\b}(x) \ra  \over \delta g_{\lambda\rho}(x')}  
=\Big( g^{\m\lambda} \langle C^{\rho\a}(x)\rangle 
+ g^{\m\rho} \langle C^{\lambda\a }(x)\rangle \nn\\
&&~~~~~~~~~~~~~~~~ -g^{\m\a} \langle C^{\lambda\rho }(x)\rangle \Big)\del_\a  \delta(x-x') 
+ {\cal O} \big(\Gamma\big)
 \ ,
\ee  
where 
$\delta \Gamma^\mu_{\a\b}
= - g^{\m\g} \delta g_{\g\sigma} \Gamma^\sigma_{\a\b} 
+{g^{\mu\g}\over 2}\big(\del_\a \delta g_{\g\b}+\del_\b \delta g_{\g\a}-\del_\g\delta g_{\a\b}  \big)$.  
The ${\cal O} \big(\Gamma\big)$ part vanishes in the flat-space limit. 
The above expressions lead to
\be
\label{master}
&& i    [ T^{\mu 0 } (x), T^{\lambda\rho} (x')] \delta(x^0-x'^{0})
= -  2     \del_\nu   {\delta   C^{\mu\nu} (x) \over  \delta g_{\lambda\rho}(x')} \Big|_{\rm{flat}} 
\nn\\
&&-\Big( \eta^{\m\lambda} T^{\rho\a}(x) 
+ \eta^{\m\rho} T^{\lambda\a }(x) 
-\eta^{\m\a} T^{\lambda\rho }(x) \Big)\del_\a  \delta(x-x')\ . 
\ee 
We have used $\del_\nu C^{\mu\nu} (x)={\cal O} \big(\Gamma\big)$, 
$\langle T^{\mu\nu} (x) T^{\lambda\rho} (x')\ra_+= \langle T^{\mu\nu} (x) T^{\lambda\rho} (x') \ra  \theta(x^0-x'^{0}) 
+ \langle T^{\lambda\rho} (x') T^{\mu\nu} (x) \ra \theta(x'^{0}-x^0)$, and 
$\del_\nu \theta(x^0-x'^{0})=\delta^0_\nu \delta(x^0-x'^{0})$.
Note the equal-time commutator should not have an explicit time-derivative of $\delta(x^0-x'^{0})$, but 
  the right-hand side of \eqref{master} presently looks like it has such a dependence. 
Consistency requires that the object $\del_\nu  {\delta   C^{\mu\nu} (x) \over  \delta g_{\lambda\rho}(x')}$ provides a cancellation. 

It can be instructive to derive explicitly the commutator involving 
only temporal-components; other components can be obtained in a similar manner.
From \eqref{master}, 
 \be
\label{t00t00}
 &&i    [ T^{00 } (x), T^{00} (x')]\delta(x^0-x'^{0})= 
-  2   \del_\nu   {\delta   C^{0\nu} (x) \over  \delta g_{00}(x')}\Big|_{\rm{flat}}  \nn\\
&&+\Big( 2T^{0i}(x)  \del_i
+ T^{00}(x)\del_0 \Big)  \delta(x-x')\ .
\ee  
Defining a  parametrization  function $t(x,x')$ via
\be
2     {\delta   C^{0\nu} (x) \over  \delta g_{00}(x')}\Big|_{\rm{flat}}  
= t^{0 \n, 0 0}(x, x') - \eta^{\n 0} T^{00}(x) \delta(x-x') \ ,
\ee 
the right-hand side of \eqref{t00t00} can be written as
 \be
 - \del_\nu  t^{0 \n, 0 0}(x, x') - \del_0 T^{00}(x) \delta(x-x') + 2T^{0 i }(x)  \del_i \delta(x-x')  \nn
\ee 
with no time-derivative on $\delta(x^0-x'^{0})$ left.
A direct manipulation using
$T^{0i}(x') \del_i \delta(x-x')=- \del_0  T^{00}(x) \delta(x-x')+ T^{0i}(x) \del_i  \delta(x-x')$ 
gives
\be
 && i    [ T^{00 } (x), T^{00} (x')] \delta(x^0-x'^{0})\nn\\
&&= \Big(T^{0i}(x)+T^{0i}(x') \Big)\del_i \delta(x-x') - \del_\nu  t^{0 \n, 0 0}(x, x')  \ .
\ee  
Similarly, using \eqref{master}, it is straightforward to consider other components.
We now tabulate the various commutators:
\be
\label{s1}
&& i    [ T^{00 } (x), T^{00} (x')] \delta_t= \Big(T^{0i}(x)+T^{0i}(x') \Big)\del_i \delta(x-x') \nn\\
&&~~~~~~~~~~~~~~~~~~~~~~~~~~~~~~ - s^{0,00}(x,x')  \  , \\
\label{s2}
&& i    [ T^{0 0 } (x), T^{0 i} (x')] \delta_t=\Big( T^{i j}(x) +T^{00}(x') \delta^{ij} \Big) \del_j \delta(x-x') \nn\\
&&~~~~~~~~~~~~~~~~~~~~~~~~~~~~~~-   s^{0,0i}(x,x')  \ ,\\
&& i    [ T^{0 0 } (x), T^{ij} (x')] \delta_t=\Big(T^{0i}(x') \del^j+ T^{0j}(x') \del^i  \\
&&~~~~~~~~~~~~~~~~~~~~~~~~~~ -\del^0  T^{ij}(x) \Big) \delta(x-x')-   s^{0,ij}(x,x')  \ ,\nn\\
\label{s3}
&& i    [ T^{0 i} (x), T^{0j} (x')]\delta_t= \Big( T^{0j}(x) \del^i+T^{0i}(x')\del^j\Big)  \delta(x-x')\nn\\
&&~~~~~~~~~~~~~~~~~~~~~~~~~~~~~~ - s^{i,0j}(x,x')  \ , \\
\label{s5}
&& i    [ T^{0 i} (x), T^{jk} (x')] \delta_t= \Big( \delta^{im} T^{jk }(x) -\delta^{ij} T^{ k m}(x')  \\
&&~~~~~~~~~~~~~~~~~~~ - \delta^{ik} T^{jm }(x') \Big)\del_m \delta(x-x') -s^{i,jk}(x,x')  \ ,\nn
\ee
where $\delta_t \equiv \delta(x^0-x'^{0})$ 
and the Schwinger term is $s^{\mu,\lambda\rho}\equiv \del_\nu  t^{\mu \n, \lambda\rho}$ with  
\be
\label{t1}
&&t^{0 \n, 0 0} = 2     {\delta   C^{0\nu} (x) \over  \delta g_{00}(x')}\Big|_{\rm{flat}}  +\eta^{\n 0} T^{00}(x) \delta(x-x') \ , \\
&&t^{0 \n, 0 i}=2   {\delta   C^{0\nu} (x) \over  \delta g_{0i}(x')}\Big|_{\rm{flat}} + \eta^{\n i} T^{00}(x) \delta(x-x') 
 \ , \\
&&t^{0 \n, ij} = 2     {\delta   C^{0\nu} (x) \over  \delta g_{ij}(x')}\Big|_{\rm{flat}}  +\Big(\eta^{\n j} T^{0i}(x) \nn\\
&&~~~~~~~~~~~~~~ +\eta^{\n i} T^{0j}(x)-\eta^{0 \n} T^{ij}(x)\Big)\delta(x-x')\ , \\
\label{t4}
&& t^{i\nu, 0 j}=2{\delta   C^{i\nu} (x) \over  \delta g_{0j}(x')}\Big|_{\rm{flat}}  \nn\\
&&~~~~~~~~~~~~~ +\Big( T^{0i}(x) \eta^{\nu j}+ T^{0\nu}(x) \delta^{ij}  \Big) \delta(x-x') \ , \\
\label{t5}
&&t^{i\nu,  jk}=2 {\delta   C^{i \nu} (x) \over  \delta g_{jk}(x')}\Big|_{\rm{flat}} \nn\\
&&~~~~~ -\Big( \delta^{ij} \eta^{\nu 0}T^{ k 0 }(x) +\delta^{ik} \eta^{\nu 0}T^{ j 0 }(x)  \Big)\delta(x-x') \ .
\ee    
The commutation relation involving purely spatial-components 
does not admit a model-independent form. 

Some constraints on the Schwinger term must be imposed 
 so that the above commutators remain consistent with the Poincar\'e relations
\be
&&i [T^{\mu\nu}(x), P^\lambda]
=\del^{\lambda} T^{\mu\nu}(x) \ ,\\
&&i [T^{\mu\nu}(x), J^{\lambda\sigma}]
=(x^\lambda \del^\sigma-x^\sigma \del^{\lambda}) T^{\mu\nu}(x)\\
&& + \eta^{\mu\lambda}T^{\sigma\nu}(x)+ \eta^{\nu\lambda}T^{\sigma\mu}(x) - \eta^{\mu\sigma}T^{\lambda\nu}(x)-\eta^{\nu\sigma}T^{\lambda\mu}(x) \ , \nn 
\ee where 
$P^\lambda= \int d{\vec x}~T^{0\mu}$, $J^{\lambda\sigma}=\int d{\vec x}~(x^\lambda T^{0\sigma}-x^\sigma T^{0\lambda})$. 
 One thus requires 
$\int d{\vec x}~s^{0,00}=\int d{\vec x}~x^i s^{0,00}=0$.      
On the other hand, one may adopt 
$\bar T_{\mu\nu}= T_{\mu\nu} - \la T_{\mu\nu} \rangle$, 
with $\la T_{\mu\nu} \rangle\sim \eta_{\mu\nu}$, and check that the above structures remain formally the same. 
Below, we shall focus on flat-space CFT with a traceless stress tensor. As 
the expectation value of a CFT stress tensor in flat-space limit is zero, we avoid additional notation $\bar T_{\mu\nu}$.

\vskip 0.1in
\noindent
{\it Effective Lightcone Commutator in CFT.---}
We now discuss in what sense a lightcone limit is taken and what the dominating structure is.  
The tracelessness condition will be imposed. 
Let
\be
ds^2= - dx^+ dx^- + \delta_{ab} d x^a dx^b \ ,
\ee 
where $(x^+, x^-)= (t+y ,t-y)$. One has
\be
\label{T++}
&&T^{\pm\pm}= -2 \big( T^0_0\mp T^0_1\big) - T^a_a \ , ~~~T^{-+}= T^a_a \ .
\ee 
In $d=2$,  $T^{++}/T^{--}$ is independent of $x^+/ x^-$, respectively, and $T^a_a=0$.
Going beyond $d=2$, we consider the following lightcone limit: 
\be
\label{LClimit}
x^- \to 0  ~~~~ {\rm with}~ x^+~{\rm fixed} \ ,
\ee   
(One can also consider $x^+ \to 0$ with $x^-$ fixed.) 
and focus on the $\it{effective}$ commutation relation, where 
stress tensors are in a correlation function.  
The reason to adopt such a scenario is that, for many purposes, such 
as in the conformal block decomposition, stress tensors always appear in a correlator
 and thus having an effective commutator would be sufficient.  

We shall focus on stress tensors with indices uncontracted, corresponding 
to the lowest-twist or largest-spin limit. 
Intuitively, the lightcone-component  $T^{++}$ should dominate near the lightcone.   
However, in $d>2$, the existence of $T^a_a$ in \eqref{T++}  
causes trouble: since purely spatial-components of the stress-tensor commutator
do not have a model-independent expression, there is no universal way to compute 
$[T^{++}(x),T^{++}(x')]$ in $d>2$. 
This obstacle may be circumvented by 
adopting an effective lightcone commutator.  
Let us here demonstrate via an example.
Consider the following CFT correlator \cite{Cardy1987}:  
\be
&&\langle T_{\mu\nu}(x_1) \phi(x_2) \phi(0) \ra
\sim a {I_\mu^\lambda(x_1) I^\sigma_\nu(x_1) A_{\lambda\sigma}(x_2) \over x_1^{2d} x_2^{2\Delta_{\phi}+2-d}}  \ , \\
&&I_\mu^\lambda(s)= \delta_\mu^\lambda - {2s^\lambda s_\mu \over s^2}\ , 
~~ A_{\lambda\sigma}(s)= {s_\lambda s_\sigma- {s^2 \over d} \eta_{\lambda\sigma}} \ ,
\ee 
where we focus on the short-distance (small $x_2$) behavior; $a$ is 
a constant and $\Delta_{\phi}$ the dimension of $\phi$. 
We will put scalar fields on a $x^+ -x^-$ plane and consider the light-cone limit $x^- \to 0$. 
The stress tensors generally allow the transverse-coordinate dependence and  we consider that stress tensors also approach to the lightcone. 
One finds the following limiting behavior under \eqref{LClimit}: 
\be
&&x_1^{2d} x_2^{2\Delta_{\phi}+2-d} \langle (T^0_0- T^0_1)(x_1) \phi(x_2) \phi(0) \ra \sim  (x^+_2)^2 \ , 
\ee 
while having $T^0_0+ T^0_1, T^{a}_a, T^{+a}$, or  $T^{-a}$ 
 in the correlator does not contribute in the same  limit.
Although this example involves only a stress tensor, one may consider 
 more stress tensors and verify that the contributions near 
the lightcone are dominated by $(T^0_0- T^0_1)^n$ with 
$n$ stress tensors. 
We thus focus on the commutator involving $T^0_0- T^0_1$ as the effective lightcone description.
In what follows, we let
\be
\widetilde T^{++}= -2 \big( T^0_0- T^0_1\big)  \ .
\ee 
 Let us also remark that the operator $\del_+ \widetilde T^{++}$ 
does not contribute in the same lightcone limit either.

We next compute the commutator of $\widetilde T^{++}$  
using \eqref{s1}, \eqref{s2}, and \eqref{s3}, together with 
 the tracelessness condition.
We find, before imposing the lightcone limit, the following 
intermediate expression: 
\be
\label{maine}
&&- i  [\widetilde T^{++}(x), \widetilde T^{++}(x') ] \nn\\
&=& 
 -4\Big(\widetilde T^{++}(x)+ \widetilde T^{++}(x') \Big) (\del_+ - \del_-) \delta^{d-1}\nn\\
&&+4\Big( T^{a}_a(x)+T^{a}_a(x')  \Big) (\del_+ - \del_-) \delta^{d-1} \nn\\
&& -4 \Big(T^{+a}(x) +T^{+ a}(x')\Big) \del_a \delta^{d-1} + \tilde s(x,x') \ ,
\ee
where $\delta^{d-1}= \delta(y- y') \delta^{d-2}(x^a-x'^a)$ and 
\be
\label{schw}
&& \tilde s(x,x')=4 \Big( \del_\nu  t^{0 \n, 0 0}(x,x') 
+   \del_\nu  t^{0 \n, 0 1}(x,x') \nn\\
&& ~~~~~~~~~~~~~~~~~~ -   \del'_\nu  t^{0 \n, 0 1}(x',x)
+  \del_\nu t^{1\nu, 0 1}(x,x') \Big) 
\ee 
 is the  corresponding Schwinger term.  
At equal time, one may write the difference $(y- y')$ as either $(x^+- x'^+)$ or $-(x^- - x'^-)$.
We have here explicitly indicated the dimensionality of the delta function.
 
As a check on \eqref{maine}, let us take $d=2$, where
 $T^{a}_a=T^{+a}=\del_+  \widetilde  T^{++}=0$. 
We have 
\be
\label{2dvv}
&&-i[\widetilde T^{++}(x^-), \widetilde T^{++}(x'^-) ]\nn\\ 
&&= 4 \big(\widetilde T^{++}(x^-)+\widetilde T^{++}(x'^-) \big)\del_{-} \delta(x^-- x'^-) +\tilde s|_{d=2} \ , 
\ee 
which is precisely the Virasoro algebra \eqref{V1}, as it must be, provided 
that the Schwinger term is related to the central charge  via
\be
\label{2dS}
\tilde s|_{d=2} = - {2c\over 3 \pi}\del^3_- \delta(x^--x'^-) \ .
\ee   
We leave the discussion on the Schwinger term to the next section.  

In $d>2$, we shall focus on the lightcone limit where
the additional  $T^{a}_a$, $T^{+a}$  pieces in \eqref{maine} are suppressed. 
The effective lightcone commutator may be written as
\be
\label{lightconeviraso}
&& -i[\widetilde T^{++}(x^+, x^a), \widetilde T^{++}(x'^+, x'^a) ]\nn\\ 
&=& -4 \Big(\widetilde T^{++}(x^+, x^a)+\widetilde T^{++}(x'^+, x'^a) \Big)\del_{+}\delta^{d-1} \nn\\
&&~~~~~~~~~~~~~~~~~~~~~~~~~~ +\tilde s(x^+,  x^a, x'^+, x'^a)\ ,
\ee   
where $\delta^{d-1}=\delta(x^+ - x'^+) \delta^{d-2}(x^a-x'^a)$. 
We have dropped the $x^-, x'^-$ dependence 
in the stress tensors and the Schwinger term to indicate that  
these contributions are localized on the lightcone. 
The lightcone limit is imposed after computing the commutation relation.   
In $d=2$, it is not necessary to impose a lightcone limit.

\vskip 0.1in
\noindent
{\it Remarks on the Schwinger Term.---}
The connection between the Schwinger term 
and $C_T$ can be 
deduced from the spectral representation \cite{CFL1991,Osborn:1999az}. 
First consider the K\"allen-Lehmann spectral form of the stress-tensor 
two-point function in unitary QFT (in Euclidean signature)
\be
\label{generaltt}
&&\langle T_{\mu\nu}(x) T_{\lambda\rho}(0)\ra\\
&&= {N_d} \int_0^\infty d\mu~  \Big(\rho^{(0)}(\mu)\Pi^{(0)}_{\mu\nu,\lambda\rho} 
+\rho^{(2)}(\mu)\Pi^{(2)}_{\mu\nu,\lambda\rho} \Big)G(x,\mu)\ , \nn
\ee 
where 
\be
&&N_d= { 2 \pi^{d/2}\over (d-1)^2 (d+1) 2^{d-1}\Gamma(d/2)}\ , \\
&&\Pi^{(0)}_{\mu\nu,\lambda\rho}= {1\over \Gamma(d)} S_{\mu\nu} S_{\lambda\rho}  \ , ~S_{\mu\nu} = \del_\mu \del_\nu-\delta_{\mu\nu} \del^2\ ,  \\
&&\Pi^{(2)}_{\mu\nu,\lambda\rho}= {d-1\over 2 \Gamma(d-1)} \Big(S_{\mu\lambda} S_{\nu\rho}+S_{\mu\rho} S_{\nu\lambda}-{2S_{\mu\nu} S_{\lambda\rho}\over d-1}\Big)\ ,\\
&&G= \int {d^dp\over (2\pi)^d} {e^{i px}\over p^2+\mu^2}=  \Big({\mu\over 2\pi |x|}\Big)^{{d\over 2}-1} {K_{{d\over 2}-1}(\mu |x|)\over 2\pi} \ .
\ee   
The spectral functions $\rho^{(0)}(\mu),  
\rho^{(2)}(\mu)$ represent spin-$0$ and spin-$2$ intermediate states, respectively.  
Restricting to CFT, scale invariance implies
\be
\label{forms}
\rho^{(0)}(\mu)= \rho^{(0)}\mu^{d-2} \delta(\mu) \ , ~~~~\rho^{(2)}(\mu)=\rho^{(2)} \mu^{d-3} \ . 
\ee 
These functions do not lead to an infrared singularity.
$\rho^{(0)}(\mu)$ does not contribute in $d>2$ while $\Pi^{(2)}_{\mu\nu,\lambda\rho}$ vanishes in $d=2$.  
With \eqref{forms}, one can compute \eqref{generaltt} and match with  \cite{Osborn:1993cr}
\be
\label{CTI}
&&\langle T_{\mu\nu}(x) T_{\lambda\rho}(0) \ra = C_T {{\cal I}_{\mu\nu,\lambda\rho}(x)\over x^{2d}} \ ,  \\
\label{defineI}
&&{\cal I}_{\mu \nu,\lambda \rho} (x) 
= {1\over 2}\Big( I_{\mu \lambda}(x) I_{\nu \rho}(x) 
+ I_{\mu \rho}(x) I_{\nu \lambda} (x) \Big) - {1\over d} \delta_{\mu\nu} \delta_{\lambda\rho}   \ , \nn
\ee 
and find, in CFTs,  
\be
\label{r2}
\rho^{(0)}= { C_T\over 2 } 
\ , ~~~~\rho^{(2)} = {d-1\over d} C_T \ .
\ee  

It is more involved to obtain the 
delta-function distribution from the Schwinger term, but it  
has been worked out a long time ago for unitary QFT \cite{RB68, KM}.
As a concrete example in higher dimensions, we focus on $d=4$ in the following. 
The relevant results in $d=4$ Lorentzian CFT, where $\rho^{(0)}$ 
does not contribute and $\rho^{(2)}= {3 C_T\over 4}$,  
are 
\be
&&F_{\mu\nu,\lambda\rho}(x-x')\equiv \la [T_{\mu\nu}(x), T_{\lambda\rho}(x')] \rangle\ ,\nn\\
&&F_{00,00}=F_{0i,0j}=F_{00,ij}= 0 \ , \nn\\
&&F_{00,0i}= -i {N_4} \int^\Lambda_0~ d\mu \rho^{(2)}(\mu) \del_i \Delta \delta^3(x-x')\nn\\
&&~~~~~~~~~~~ - i {N_4 \over 2} \rho^{(2)}  \del_i \Delta^2 \delta^3(x-x')\nn\\
&&~~~~~~~= -i {C_T \pi^2 \over 480} \Big(\Lambda^2  + \Delta \Big) \del_i \Delta\delta^3(x-x') \ , 
\ee 
where $\Delta$ is a Laplacian.
Only the equal-time commutators with an odd number of temporal indices 
are non-zero since the causal propagator is odd in time; 
$F_{0i,jk}$ can be non-zero, but these 
components are irrelevant in the lightcone limit. 
Note that the fifth-derivative ``boundary"  
term generally exists in $d=4$, unless 
one has restricted to field theories with 
$\rho^{(2)}(\mu \to \infty)=0$, which is however not a CFT.

In $d=2$, a similar analysis gives the following 
non-zero contribution:
\be
F_{00,01}&=&  -i {\pi C_T\over 6}   \del^3_1 \delta(x^1-x'^1) \ ,
\ee 
where we have recalled \eqref{forms} and \eqref{r2}, and
this gives 
\be
\langle \tilde s \ra|_{d=2}= - {4 \pi C_T\over 3}   \del^3_- \delta(x^--x'^-) \ , 
\ee 
which reproduces the required identification \eqref{2dS} in $d=2$ CFT, with  
$C_T= {c\over 2 \pi^2}$ where $c=1$ for a free boson.  

We remark that, a prior, the Schwinger term could be a $q$-number in general QFT. 
From the Virasoro algebra, one knows the corresponding Schwinger term must be a $c$-number in $d=2$ CFT with $\tilde s|_{d=2}=\langle \tilde s \ra|_{d=2}$.
For $d>2$,  we shall here assume that we are restricted in the class of CFTs where the Schwinger term is a $c$-number, at least in the lightcone limit. 
A more general question whether the Schwinger term might always be a $c$-number goes beyond the scope of the present work and will not be addressed here.  
Note the Schwinger term must have a consistent dimension and the requirement that the $d\to 2$ limit of \eqref{maine} reproduces the Virasoro algebra implies 
the Schwinger term in general $d$ should not touch the coefficient of the existing $\widetilde T^{++}$ piece.  
(One may adopt additional limits, such as a large $C_T$, to suppress possible unwanted contributions, but we do not consider such a limit here.)

On the other hand, a direct canonical computation  shows that   the 
Schwinger terms \eqref{t1}-\eqref{t4} in $d>2$ free theories 
simply vanish, as first pointed out in \cite{BD1967}, while \eqref{t5}, which is 
however irrelevant in the lightcone limit, can be non-zero but its expression is model-dependent.  
(Related computation details can be found in \cite{JT}.)  
The authors of \cite{BD1967, JT} then argued that quantum effects are responsible for producing 
the anomalous $c$-number delta-function distribution predicted from the spectral forms. 
While their results suggest that the free theories belong to the class of theories 
we are interested in, it would be interesting to revisit the free-theory calculations 
and derive the $c$-number contribution in view of the results presented here.

In the class of $d=4$ CFTs, for instance, where the 
Schwinger term is effectively a $c$-number near the lightcone,  
we may write the lightcone effective algebra as 
\be
\label{4dVira}
&& -i[\widetilde T^{++}(x^+, x^a), \widetilde T^{++}(x'^+, x'^a)]\nn\\
&=&- 4 \Big(\widetilde T^{++}(x^+, x^a)+\widetilde T^{++}(x'^+, x'^a) \Big)\del_+\delta^{3} \\
&&~~~~~~~~~~~~~~~~~~~ + {C_T\pi^2 \over 60} \Big(\Lambda^2  +\Delta \Big)\del_+ \Delta\delta^3\ , \nn
\ee   
where $\delta^{3}= \delta(x^+ - x'^+) \delta^{2}(x^a-x'^a)$. 
The appearance of the UV divergence in the Schwinger term is expected to be 
a general figure in $d>2$ CFTs, based on essentially dimensional analysis.  
The coefficient of the power-law divergence has no universal meaning and thus  
we shall focus on the universal finite piece 
(the coefficient of the highest-order derivative of the delta function) in the Schwinger term  
and subtract the divergence off  when computing a correlator.   
 
We hope to discuss the applications of the lightcone stress-tensor commutation relation elsewhere.   
In particular, it would be interesting to realize the lightcone algebra in a holographic framework and 
explore potential connections 
with the lowest-twist universality \cite{Fitzpatrick:2019zqz} 
and also some recent works \cite{Casini:2017roe, Afkhami-Jeddi:2017rmx, Cordova:2018ygx, Belin:2019mnx, Kologlu:2019mfz, Balakrishnan:2019gxl}. 
\\

I thank L.~Fitzpatrick, T.~Hartman, C.~Herzog, K.~Jensen, H.~Osborn, and A.~Tseytlin 
for discussions and useful comments.
This work was supported in part by the U.S.
Department of Energy Office of Science under award number DE-SC0015845 
and in part by the Simons Collaboration grant on the Non-Perturbative Bootstrap.

\end{document}